\documentclass[journal,twoside]{IEEEtran}
\usepackage{graphicx}
\usepackage{cite}
\usepackage{amssymb,amsmath}
\usepackage{subfigure}
\usepackage{amssymb,amsmath}
\usepackage{subfigure}
\usepackage[T1]{fontenc}
\usepackage[latin9]{inputenc}
\usepackage{float}
\usepackage[english]{babel}
\usepackage{epstopdf}
\usepackage{etoolbox}
\usepackage{algorithm}
\usepackage{algorithmic}

\preto\subequations{\ifhmode\unskip\fi}
\makeatletter
\newcommand\restartchapters{\par
  \setcounter{chapter}{0}%
  \setcounter{section}{0}%
  \gdef\@chapapp{\chaptername}%
  \gdef\thechapter{\@arabic\c@chapter}}
\makeatother

\newtheorem{lemma}{\bf Lemma}

\newcommand{\st}{\mathrm{s.\,t.} }

\DeclareMathOperator{\tr}{tr}

\begin{document}

\title{ Robust Beamforming for Secrecy Rate in Cooperative Cognitive Radio Multicast Communications}
\author{
\IEEEauthorblockN{Van-Dinh Nguyen$^{\dag}$, Trung Q. Duong$^{\S}$, Oh-Soon Shin$^{\dag}$,  Arumugam Nallanathan$^*$,  and 	George K. Karagiannidis$^{\P}$   } \\

\IEEEauthorblockA{$^{\dag}$School of Electronic Engineering $\&$ Department of  ICMC Convergence Technology, Soongsil University, Korea.	(corresponding author, e-mail: osshin@ssu.ac.kr)\\
                    $^{\S}$School of Electronics, Electrical Engineering and Computer Science, Queen's University Belfast, UK\\
										$^*$Center for Telecommunications Research, King's College London, U.K\\
										$^{\P}$Department of Electrical and Computer Engineering, Aristotle University of Thessaloniki, Greece
							}}
\maketitle
\thispagestyle{empty}
\pagestyle{empty} 
\vspace{-1.8cm}
\begin{abstract}
  In this paper, we propose a cooperative approach to improve the security of both primary and secondary systems in cognitive radio multicast communications. During their access to the frequency spectrum licensed to the primary  users, the secondary unlicensed users  assist the primary system in fortifying security by sending a jamming noise to the eavesdroppers, while  simultaneously protect themselves from eavesdropping.  The main objective of this work is to maximize the secrecy rate of the secondary system, while  adhering to all individual primary users' secrecy rate constraints.	In the case of passive eavesdroppers and imperfect channel state information  knowledge at the transceivers,  the utility function of interest is nonconcave and involved constraints are nonconvex, and thus, the optimal solutions are troublesome.  To address this problem, we propose an iterative algorithm to arrive at a local optimum of the considered problem. The proposed iterative algorithm is guaranteed to achieve a Karush-Kuhn-Tucker solution. 
\end{abstract}

\section{Introduction} \label{Introduction}

Recently,  physical  layer (PHY) security for wireless communications has become  an important research area. The underlying idea is to guarantee a positive secrecy rate of legitimate users by exploiting the random characteristics of the wireless channel. In particular,  the authors in  \cite{Gopala} proposed a low-complexity on/off power allocation strategy to attain secrecy  under the assumption of full channel state information (CSI). The use of cooperative jamming noise (JN) was proposed in  \cite{Tekin}, where users who are prevented from transmitting according to a certain policy will block the eavesdropper and thereby assist the remaining users. From a  quality-of-service perspective, a secret transmit beamforming approach  was considered in  \cite{Mukherjee_1}, in order to predetermine the signal-to-interference-plus-noise-ratio (SINR) target at the destination and/or at the eavesdropper.

Being a critical issue, PHY security of cognitive radio networks (CRNs), which are faced with specific security risks due to the broadcasting nature of  radio signals \cite{Nguyen:COMML16,Yeoh:TVT15,Lui:TCOM16}, however,  has not been well investigated until  recently, e.g., in  \cite{Nguyen:TIFS:16,Pei,Nguyen,Nguyen_15,Zhu:VT:15}. More specifically, in \cite{Nguyen:TIFS:16} and \cite{Pei}, multi-antennas at the secondary transmitter  were utilized to attain  beamforming that maximizes the secrecy capacity of the secondary system,  while adhering to the peak interference constraint at the primary receiver.  Furthermore, a simple case with  single antenna at the eavesdropper was considered in \cite{Nguyen}.  In \cite{Zhu:VT:15}, the authors considered a CRN model, where both the primary user (PU) and the secondary user (SU) send their confidential messages to intended receivers that are surrounded by a single eavesdropper.  

In this paper, we consider the PHY security in cooperative cognitive radio multicast communications, where the eavesdroppers intend to wiretap  data from both the primary  and  secondary systems. We assume that the primary transmitter  is equipped only with a single antenna, which implies that the primary transmitter cannot generate a jamming signal or design a beamforming vector to protect itself from the eavesdroppers. The secrecy capacity of the primary system is improved by implementing a cooperative framework between the primary  and secondary systems. Specifically, the primary  allows the secondary system to share its spectrum, and in return the secondary system sends jamming noise to degrade the eavesdropper's channel, in order to protect the primary system.  
 Specifically, the main contributions of this paper can be summarized as follows:
\begin{itemize}
	
	\item We design a joint information and jamming signal at the secondary transmitter, where information is intended for secondary receivers  and jamming noise is intended for eavesdroppers. The main objective is to maximize the secrecy rate of the secondary system, while satisfying the minimum secrecy rate requirement for each legitimate user of the primary system as well as  the  power constraint.
	\item We propose a method to find the approximate solution for optimal transmit beamforming,  by providing the convexity of the problem that is considered through the use of   a convex approximation.  The optimal solutions of transmit beamforming for the confidential information and jamming noise do not fix the transmit strategy. 
	\item We provide extensive numerical results to justify the novelty of the proposed algorithm and compare its performance with the known solutions. In particular, the numerical results demonstrate fast convergence of the proposed algorithm and significantly improve the secrecy rate compared with the known solutions. We should remark that our results are more general than in \cite{Zhu:VT:15}, which was considered under the assumptions of one eavesdropper and perfect CSI.
	
	\end{itemize}

\section{System Model and Optimization Problem} \label{System Model}

\begin{figure}[t]
\centering
\includegraphics[trim=0.0cm -0.1cm 0.0cm 0.0cm, width=0.39\textwidth]{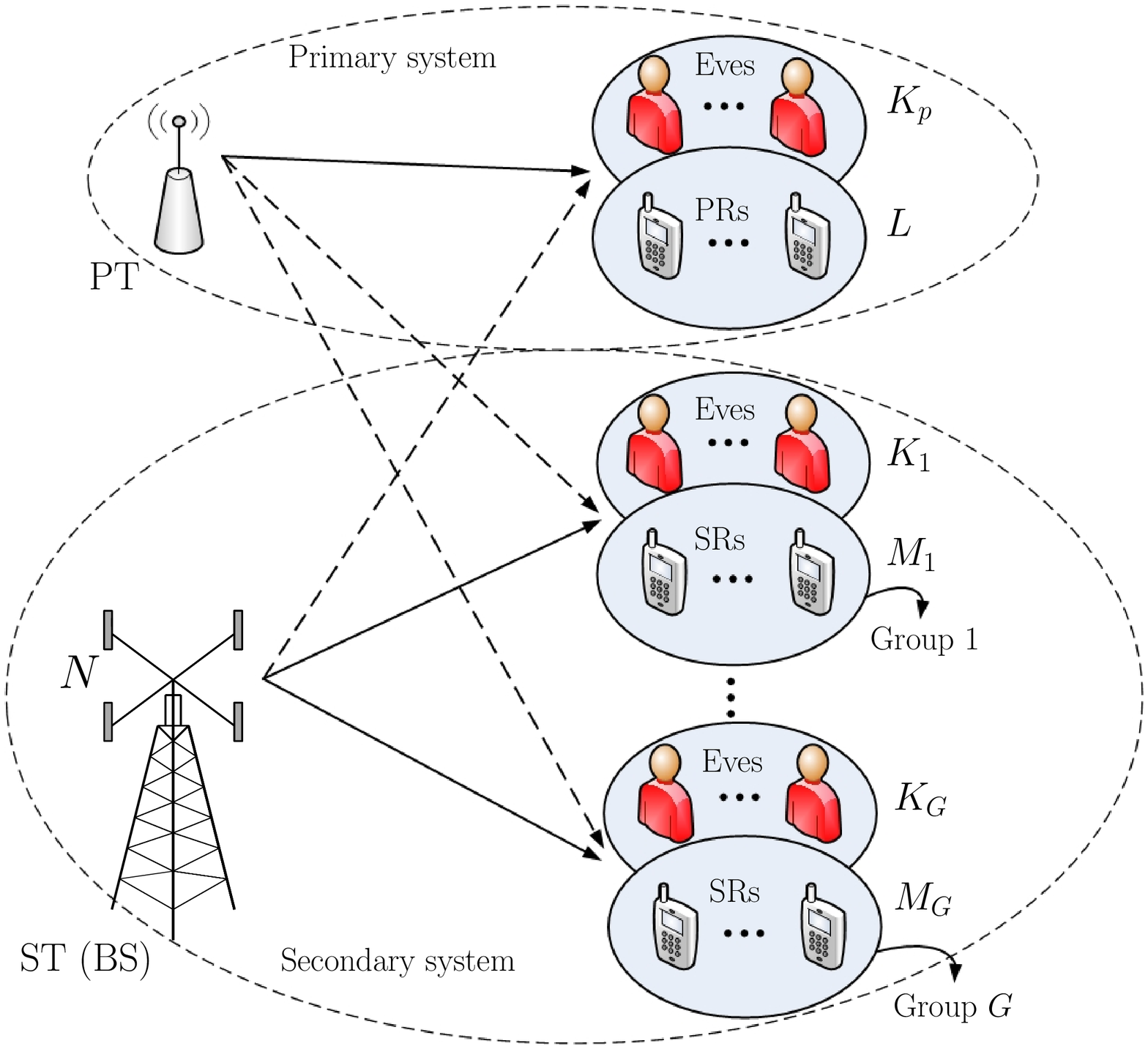}
\caption{A cooperative CRN multicast transmission  model with multiple eavesdroppers.}
\label{fig:zoneselect}
\end{figure}
\subsection{System  Model} \label{Cognitive Radio Network}
 The primary system consists of one primary transmitter (PT) and $L$ primary receivers (PRs), while the secondary system consists of one secondary transmitter  (ST) and $M$ secondary receivers (SRs),  as illustrated in Fig. 1.   The ST, which is a base station (BS),  is equipped with $N$ antennas,  whereas all  other nodes are equipped with only one antenna.  The opportunistic spectrum access is improved by assigning the ST  to send  $G$ information bearing signals $s_g, g=1, \cdots, G$, where $s_g$ is the information being sent to the $g$-th group  with unit average power $\mathbb{E}\{\left|s_g\right|^2\}=1$.  We assume that each individual multicast group $\mathcal{G}_g$ in the secondary system consists of $M_g$ the secondary receivers. Specifically, the number of  SRs in  group $\mathcal{G}_g$ is denoted by $\mathcal{S}_g=\left\{1,\cdots, m_g,\cdots, M_g\right\}$. Then, the total number of SRs  in the secondary system with multicast transmission is indeed $M=\sum_{g=1}^GM_g$.  
Regarding security, we assume that the eavesdroppers (Eves) potentially intend to wiretap and decode confidential messages from both the primary  and  secondary systems \cite{Mokari}.  We assume that each group $\mathcal{G}_g$ and the PRs are respectively wiretapped by a set of Eves such as $\mathcal{K}_{e,g}\triangleq\left\{1,\cdots,k_g,\cdots,K_g\right\}$, $\forall g$ and $\mathcal{K}_{p}\triangleq\left\{1,\cdots,k_p,\cdots,K_p\right\}$. This implies that at the same time, each legitimate user is wiretapped by a separate  group of Eves.

We aim to design multiple beamforming vectors at  the ST, one for the JN and the other  for its own information signal, to protect both the primary and secondary systems.   The transmit power at the PT is $P_p >0$ and the data intended for the PR is $x_p$ with unit average power $\mathbb{E}\{\left|x_p\right|^2\}=1$. Before transmission, the data of the SRs $s_g$ in the group $\mathcal{G}_g$ is weighted to the $N\times 1$ beamforming vector $\mathbf{w}_g$, $\forall g$.  Hence, all the transmitted signals at the ST can be expressed through a vector $\mathbf{x}_s$ as  
\begin{equation}
\mathbf{x}_s = \sum_{g=1}^G\mathbf{w}_gs_g+\mathbf{u}
\label{eq:xs}
\end{equation}
where  $\mathbf{u}$ is the artificial noise vector, whose elements are zero-mean complex Gaussian random variables with covariance matrix $\mathbf{U}\mathbf{U}^{H}$, such that $\mathbf{u}\sim	\mathcal{CN}(\mathbf{0}, \mathbf{U}\mathbf{U}^H)$, where $\mathbf{U}\in\mathbb{C}^{N\times N}$. The artificial noise  $\mathbf{u}$ is assumed to be unknown to all SRs, PRs, and Eves. For notational simplicity, we define $\mathbf{w}\triangleq[\mathbf{w}_{1}^T, \mathbf{w}_{2}^T,\cdots,\mathbf{w}_{G}^T]^T\in\mathbb{C}^{NG\times 1}$.

The corresponding SINR at the $l$-th $\mbox{PR}$ for $l=1,\cdots, L$ and the $k_p$-th $\mbox{Eve}$  for $k_p=1,\cdots, K_p$  are respectively given by\footnote{$\|\cdot\|$ and $|\cdot|$ denote the Euclidean norm of a matrix or vector and  the magnitude  of a complex scalar, respectively.}
\begin{IEEEeqnarray}{rCl}
\Gamma_{p,l}(\mathbf{w},\mathbf{U})&=&\frac{P_p|h_l|^2}{\sum_{g=1}^G|\mathbf{f}_l^{H}\mathbf{w}_g|^2+ \|\mathbf{f}_l^{H}\mathbf{U}\|^2+\sigma_l^2}\label{eq:SINR:pr}\\
\Gamma_{e,k_p}(\mathbf{w},\mathbf{U})&=&\frac{P_p|g_{k_p}|^2}{\sum_{g=1}^G|\mathbf{f}_{k_p}^{H}\mathbf{w}_g|^2+ \|\mathbf{f}_{k_p}^{H}\mathbf{U}\|^2+\sigma_{k_p}^2}\label{eq:SINR:pre}
\end{IEEEeqnarray} 
where $h_{l}\in\mathbb{C}$, $g_{k_p} \in\mathbb{C}$, $\mathbf{f}_{l} \in\mathbb{C}^{N\times 1}$, and $\mathbf{f}_{k_p}\in\mathbb{C}^{N\times 1}$ are  the respective baseband equivalent channels of the links PT $\rightarrow$ $l$-th $\mbox{PR}$, PT $\rightarrow$ $k_p$-th $\mbox{Eve}$, ST $\rightarrow$ $l$-th $\mbox{PR}$, and ST $\rightarrow$ $k_p$-th $\mbox{Eve}$. $\sigma_l^2$ and $\sigma_{k_p}^2$ are the variance of the additive white Gaussian noise (AWGN) at the $l$-th $\mbox{PR}$  and  ${k_p}$-th $\mbox{Eve}$, respectively.

The respective SINR at the $m_g$-th $\mbox{SR}$  in the group $\mathcal{G}_g$ and the $k_g$-th  $\mbox{Eve}$  are given by
\begin{IEEEeqnarray}{rCl}\label{eq:SINR_ekg}
&&\Gamma_{s,m_g}(\mathbf{w},\mathbf{U})=\nonumber\\
&&\frac{|\mathbf{h}_{m_g}^{H}\mathbf{w}_g|^2}{\sum_{i=1, i\neq g }^G|\mathbf{h}_{m_g}^{H}\mathbf{w}_i|^2+ \|\mathbf{h}_{m_g}^{H}\mathbf{U}\|^2+P_p|f_{m_g}|^2+\sigma_{m_g}^2}\label{eq:SINR:sr}\\
&&\Gamma_{e,k_g}(\mathbf{w},\mathbf{U})=\nonumber\\
&& \frac{|\mathbf{g}_{k_g}^{H}\mathbf{w}_g|^2}{\sum_{i=1,i\neq g}^G|\mathbf{g}_{k_g}^{H}\mathbf{w}_i|^2+\|\mathbf{g}_{k_g}^{H}\mathbf{U}\|^2+P_p|f_{k_g}|^2+\sigma_{k_g}^2}\label{eq:SINR:se}
\end{IEEEeqnarray} 
where $\mathbf{h}_{m_g}\in\mathbb{C}^{N\times 1}$, $\mathbf{g}_{k_g}\in\mathbb{C}^{N\times 1}$, $f_{m_g}\in\mathbb{C}$, and $f_{k_g} \in\mathbb{C}$ are the corresponding baseband equivalent channels of the links ST $\rightarrow$ $m_g$-th $\mbox{SR}$, ST $\rightarrow$ $k_g$-th $\mbox{Eve}$, PT $\rightarrow$ $m_g$-th $\mbox{SR}$, PT $\rightarrow$ $k_g$-th $\mbox{Eve}$. $\sigma_{m_g}^2$ and $\sigma_{k_g}^2$ are the variance of AWGN at the $m_g$-th $\mbox{PR}$  and $k_g$-th $\mbox{Eve}$, respectively. 

The achievable secrecy rate for the $l$-th PR  of the primary system, denoted by $C_{p,l}(\mathbf{w},\mathbf{U})$, can be expressed as \cite{Gopala}
\begin{equation}
\begin{aligned}
C_{p,l}(\mathbf{w},\mathbf{U})=&\Bigl[\log_2\bigl(1+\Gamma_{p,l}(\mathbf{w},\mathbf{U})\bigr)\\
&-\underset{k_p\in\mathcal{K}_p}{\max}\log_2\bigl(1+\Gamma_{e,k_p}(\mathbf{w},\mathbf{U})\bigr)\Bigr]^+
\label{eq:C_pr}
\end{aligned}
\end{equation}
where  $\left[x\right]^+=\max\left\{0,x\right\}$.
The achievable secrecy rate for the  $m_g$-th SR of the secondary system, denoted by $C_{s,m_g}(\mathbf{w},\mathbf{U})$, can be expressed as \cite{Gopala}
\begin{equation}
\begin{aligned}
C_{s,m_g}(\mathbf{w},\mathbf{U})=&\Bigl[\log_2\bigr(1+\Gamma_{s,m_g}(\mathbf{w},\mathbf{U})\bigr)\\
&-\underset{k_g\in\mathcal{K}_{e,g}}{\max}\log_2\bigr(1+\Gamma_{e,k_g}(\mathbf{w},\mathbf{U})\bigr)\Bigr]^+.
\label{eq:C_sk}
\end{aligned}
\end{equation}

\subsection{Optimization Problem Formulation}
The objective of the system design is to maximize the minimum (max-min) secrecy rate  of the secondary system  while satisfying  the minimum quality-of-service (QoS) requirements, such as the  secrecy rate achievable for the primary system as follows
\begin{IEEEeqnarray}{rCl}\label{eq:problem_1}
\mathbf{P.1}:\quad\underset{\mathbf{w}, \mathbf{U}}{\mathrm{\max}}\mathop{\mathrm{\min}}\limits_{m_g\in\mathcal{S}_{g},   g\in\mathcal{G}}
                   &&\quad C_{s,m_g}(\mathbf{w},\mathbf{U})\IEEEyessubnumber\label{eq:8a}\\
  \st&&\ C_{p,l}(\mathbf{w},\mathbf{U})\geq \bar{R}_{p,l},\, l\in\mathcal{L} \IEEEyessubnumber\label{eq:8b}\\
     		  && \  \sum\nolimits_{g=1}^G\|\mathbf{w}_g\|^2+ \|\mathbf{U}\|^2 \leq P_{s}\, \IEEEyessubnumber\label{eq:8c}
\end{IEEEeqnarray}
where $\mathcal{L}\triangleq\{1,\cdots, L\}$ and $\mathcal{G}\triangleq\{1,\cdots, G\}$. In \eqref{eq:8b}, $\bar{R}_{p,l} > 0$ are the minimum secrecy rate requirement for each legitimate user  of the primary system. 

\subsection{CSI Model}
We consider a realistic scenario, where the instantaneous CSI between  ST and PRs is  imperfect and  Eves are  passive. Specifically, the CSI of the link between the ST and PRs is given as \cite{Li}
\begin{equation}\begin{aligned}
&\mathbf{f}_l=\mathbf{\hat{f}}_l+\Delta\mathbf{f}_l,\; \forall l\\
&\Omega_l\triangleq\{\Delta{\mathbf{f}}_l\in \mathbb{C}^{N\times 1}:\Delta{\mathbf{f}}_l^{H}\Delta{\mathbf{f}}_l\leq \delta^2_l\}
\end{aligned}\label{eq:imperfect:channel}\end{equation}
where $\mathbf{\hat{f}}_l$ is the channel estimate of the $l$-th PR  available at the ST,  and $\Delta\mathbf{f}_l$ represents the associated CSI error. For notational simplicity, we define $\Omega_l$ as a set of all possible CSI errors associated with the $l$-th PR. We assume that $\Delta\mathbf{f}_l$ are deterministic and bounded, and therefore $ \delta_l$ represents the size of the uncertainty region of the estimated CSI of the  $l$-th PR.

For the passive Eves, we further assume that the entries of $g_{k_p}$,  $\mathbf{f}_{k_p},\;\forall k_p$, $f_{k_g}$, and  $\mathbf{g}_{k_g},\;\forall k_g$, follow independent and identically distributed  (i.i.d.) Rayleigh fading, and that  the instantaneous CSI of these wiretap channels is not available at ST. These assumptions of passive Eves are commonly used in the literature \cite{Nguyen,  Zhou, Li}. Meanwhile, the channels $\mathbf{h}_{m_g}, \forall m, g,$ are  assumed to be perfectly known since the SRs are active users in the secondary system.

\section{Proposed Solution }\label{Imperfect-CSI}


In this section, we propose an iterative algorithm that arrives a local optimum of the considered optimization problem. As the first step, we convert \eqref{eq:problem_1} to another equivalent form as
\begin{IEEEeqnarray}{lCl}\label{eq:rew:1}
&&\underset{\mathbf{w}, \mathbf{U}, \boldsymbol{t}, z}{\mathrm{\mathrm{maximize}}}\mathop{\mathrm{\min}}\limits_{m_g\in\mathcal{S}_{g},g\in\mathcal{G}}                    \left\{\log_2\bigl(1+\Gamma_{s,m_g}(\mathbf{w},\mathbf{U})\bigl) - t_{g}\right\} \IEEEyessubnumber\label{eq:rew:a}\ \ \\
  &&\st\ \log_2\bigl(1+\Gamma_{e,k_g}(\mathbf{w},\mathbf{U})\bigl) \leq t_g,\ k_g\in\mathcal{K}_{e,g}, g\in\mathcal{G} \IEEEyessubnumber\label{eq:rew:b} \\         
	&&\qquad \log_2\bigl(1+\Gamma_{p,l}(\mathbf{w},\mathbf{U})\bigr) - z\geq \bar{R}_{p,l},\ l\in\mathcal{L} \IEEEyessubnumber\label{eq:rew:c}\\
	&&  \qquad \log_2\bigl(1+\Gamma_{e,k_p}(\mathbf{w},\mathbf{U})\bigl) \leq z,\ k_p\in\mathcal{K}_p\IEEEyessubnumber\label{eq:rew:d}\\
     		  &&\qquad \eqref{eq:8c}\IEEEyessubnumber\label{eq:rew:e}
\end{IEEEeqnarray}
where $\boldsymbol{t}\triangleq \{t_{g}\}$ and $z$ are the maximum allowable rate for Eve to wiretap the information from the ST and PT, respectively.
The equivalence of \eqref{eq:problem_1} and \eqref{eq:rew:1} can be easily confirmed by justifying that  the constraint  \eqref{eq:rew:b} must hold with equality at optimum.

 Based on the above setting and the assumptions in Section II. C, the optimization problem  $\mathbf{P.1}$ can be reformulated as
\begin{IEEEeqnarray}{rCl}\label{eq:imcsi:1}
\mathbf{P.2}:&&\ \underset{\mathbf{w}, \mathbf{U}, \boldsymbol{t}, z, \varphi}{\mathrm{\mathrm{maximize}}}\quad  \varphi\IEEEyessubnumber\label{eq:imcsi:a}\\
   \st &&\ \log_2\bigl(1+\Gamma_{s,m_g}(\mathbf{w},\mathbf{U})\bigl) - t_{g} \geq  \varphi, \forall m_g, \forall g \IEEEyessubnumber\label{eq:imcsi:b} \\
	&& \max_{\mathbf{g}_{k_g}, f_{k_g}}\log_2\bigl(1+\Gamma_{e,k_g}(\mathbf{w},\mathbf{U})\bigl) \leq t_g,\ \forall k_g, \forall g \IEEEyessubnumber\label{eq:imcsi:c} \\         
	&& \min_{\Delta\mathbf{f}_l\in\Omega_l}\log_2\bigl(1+\Gamma_{p,l}(\mathbf{w},\mathbf{U})\bigl) - z\geq \bar{R}_{p,l},\ \forall l\IEEEyessubnumber\label{eq:imcsi:d}\\
	&&   \max_{\mathbf{g}_{k_p}, f_{k_p}}\log_2\bigl(1+\Gamma_{e,k_p}(\mathbf{w},\mathbf{U})\bigl) \leq z,\ \forall k_p\IEEEyessubnumber\label{eq:imcsi:e}\\
     		  &&  \eqref{eq:8c} \IEEEyessubnumber\label{eq:imcsi:f}.
\end{IEEEeqnarray}
where $\varphi$ is newly introduced variable. Observe that the objective function is monotonic in its argument, therefore, we now only deal with the nonconvex constraints \eqref{eq:imcsi:b}-\eqref{eq:imcsi:e}. Let us treat the constraint \eqref{eq:imcsi:b} first. 
As the first step, \eqref{eq:SINR:sr} is equivalently rewritten by
\begin{equation}\label{f1sinr1}
\Gamma_{s,m_g}(\mathbf{w},\mathbf{U})=\frac{\bigl|\mathbf{h}_{m_g}^{H}\mathbf{w}_g\bigr|^2}{\chi_{s,m_g}(\mathbf{w},\mathbf{U})}
\end{equation}
where
\[
\chi_{s,m_g}(\mathbf{w},\mathbf{U})=\sum_{i=1, i\neq g }^G|\mathbf{h}_{m_g}^{H}\mathbf{w}_i|^2+ \|\mathbf{h}_{m_g}^{H}\mathbf{U}\|^2+P_p|f_{m_g}|^2+\sigma_{m_g}^2.
\]
From \eqref{f1sinr1}, it follows that
\begin{equation}\label{eq:srlog:1}\begin{aligned}
&\ln\Bigr(1+\frac{|\mathbf{h}_{m_g}^{H}\mathbf{w}_g|^2}{\chi_{s,m_g}(\mathbf{w},\mathbf{U})}\Bigr)  \\
&\qquad = - \ln\Bigr(1 - \frac{|\mathbf{h}_{m_g}^{H}\mathbf{w}_g|^2}{\chi_{s,m_g}(\mathbf{w},\mathbf{U})+|\mathbf{h}_{m_g}^{H}\mathbf{w}_g|^2}\Bigr). 
\end{aligned}\end{equation}
From the fact that $0\leq\frac{|\mathbf{h}_{m_g}^{H}\mathbf{w}_g|^2}{\chi_{s,m_g}(\mathbf{w},\mathbf{U})+|\mathbf{h}_{m_g}^{H}\mathbf{w}_g|^2}\triangleq \Phi(\mathbf{w},\mathbf{U}) < 1$, the function $- \ln\bigr(1 - \Phi(\mathbf{w},\mathbf{U})\bigr)$ is jointly convex w.r.t. the involved variables \cite{Stephen}, which is useful for developing an approximate solution for \eqref{eq:srlog:1}. In particular, at feasible point $\bigl(\mathbf{w}^{(n)},\mathbf{U}^{(n)}\bigr)$, we have\footnote{Hereafter, suppose the value of $(\mathbf{w},\mathbf{U})$ at the $(n+1)$-th iteration in an iterative algorithm presented shortly is denoted by $(\mathbf{w}^{(n)},\mathbf{U}^{(n)})$.}
\begin{eqnarray}
&&- \ln\Bigr(1 - \frac{|\mathbf{h}_{m_g}^{H}\mathbf{w}_g|^2}{\chi_{s,m_g}(\mathbf{w},\mathbf{U})+|\mathbf{h}_{m_g}^{H}\mathbf{w}_g|^2}\Bigr)\geq \nonumber\\
&&- \ln\Bigr(1 - \frac{|\mathbf{h}_{m_g}^{H}\mathbf{w}_g^{(n)}|^2}{\chi_{s,m_g}(\mathbf{w}^{(n)},\mathbf{U}^{(n)})+|\mathbf{h}_{m_g}^{H}\mathbf{w}_g^{(n)}|^2}\Bigr)\nonumber\\
&&-\ \Gamma_{s,m_g}\bigr(\mathbf{w}^{(n)},\mathbf{U}^{(n)}\bigr)+2\frac{\Re\bigl\{(\mathbf{w}_g^{(n)})^H\mathbf{h}_{m_g}
\mathbf{h}_{m_g}^{H}\mathbf{w}_g\bigr\}}{\chi_{s,m_g}(\mathbf{w}^{(n)},\mathbf{U}^{(n)})}\nonumber\\
&&-\  \frac{\Gamma_{s,m_g}\bigr(\mathbf{w}^{(n)},\mathbf{U}^{(n)}\bigr)\Bigl(\chi_{s,m_g}(\mathbf{w},\mathbf{U})+|\mathbf{h}_{m_g}^{H}\mathbf{w}_g|^2\Bigr)}{
\Bigl(\chi_{s,m_g}(\mathbf{w}^{(n)},\mathbf{U}^{(n)})+|\mathbf{h}_{m_g}^{H}\mathbf{w}_g^{(n)}|^2\Bigl)}\nonumber\\
&&:= \mathcal{F}_{m_g}^{(n)}(\mathbf{w},\mathbf{U}).
\label{eq:srlog:2}
\end{eqnarray}
Note that $\mathcal{F}_{m_g}^{(n)}(\mathbf{w},\mathbf{U})$ is convex and is global lower bound of $- \ln\bigr(1 - \Phi(\mathbf{w},\mathbf{U})\bigr)$. Therefore, the following equality holds at optimum
\begin{equation}\label{eq:srlog:3}\begin{aligned}
&\mathcal{F}_{m_g}^{(n)}(\mathbf{w}^{(n)},\mathbf{U}^{(n)})=\\
&- \ln\Bigr(1 - \frac{|\mathbf{h}_{m_g}^{H}\mathbf{w}_g^{(n)}|^2}{\chi_{s,m_g}(\mathbf{w}^{(n)},\mathbf{U}^{(n)})+|\mathbf{h}_{m_g}^{H}\mathbf{w}_g^{(n)}|^2}\Bigr).
\end{aligned}\end{equation}
It implies that we can iteratively replace $- \ln\bigr(1 - \Phi(\mathbf{w},\mathbf{U})\bigr)$ by $\mathcal{F}_{m_g}^{(n)}(\mathbf{w}^{(n)},\mathbf{U}^{(n)})$ to achieve a convex approximation of \eqref{eq:imcsi:b} \cite{Marks:78}. Hence, by substituting \eqref{f1sinr1}, \eqref{eq:srlog:1}, and \eqref{eq:srlog:2} to \eqref{eq:imcsi:b}, we have 
\begin{equation}\label{eq:rew:2:b:equi}
\mathcal{F}_{m_g}^{(n)}(\mathbf{w},\mathbf{U}) \geq (\varphi + t_g)\ln2.
\end{equation}
 It is now clear that the difficulty in solving \eqref{eq:imcsi:1} is due to \eqref{eq:imcsi:c}-\eqref{eq:imcsi:e} since the remaining constraints are convex and approximate convex. Instead of this, we can find a sub-optimal solution
of \eqref{eq:imcsi:1} as follows
\begin{IEEEeqnarray}{rCl}\label{eq:imcsi:2}
&&\underset{\mathbf{w}, \mathbf{U}, \boldsymbol{t}, z, \varphi, \boldsymbol{\phi},\boldsymbol{\alpha}, \beta}{\mathrm{\mathrm{maximize}}}\quad  \varphi\IEEEyessubnumber\label{eq:imcsi:2:a}\\
   &&\st \  
	\log_2\bigl(1+\phi_g\bigl) \leq t_g,\  g\in\mathcal{G} \IEEEyessubnumber\label{eq:imcsi:2:b} \\   
			&&\	\Pr\Bigl(\underset{k_g\in\mathcal{K}_{e,g}}{\max}\ \Gamma_{e,k_g}(\mathbf{w},\mathbf{U})\leq \phi_g\Bigl) \geq \epsilon_g ,\  g\in\mathcal{G} \IEEEyessubnumber\label{eq:imcsi:2:c} \\   
	&& \ \log_2\bigl(1+\alpha_l\bigl) - z\geq \bar{R}_{p,l},\ l\in\mathcal{L} \IEEEyessubnumber\label{eq:imcsi:2:d}\\
	&& \ \min_{\Delta\mathbf{f}_l\in\Omega_l}\Gamma_{p,l}(\mathbf{w},\mathbf{U}) \geq \alpha_l,\ l\in\mathcal{L} \IEEEyessubnumber\label{eq:imcsi:2:e}\\
	&&  \ \log_2\bigl(1+\beta\bigl) \leq z\IEEEyessubnumber\label{eq:imcsi:2:f}\\
	&&  \ \Pr\Bigl(\underset{k_p\in\mathcal{K}_p}{\max}\Gamma_{e,k_p}(\mathbf{w},\mathbf{U}) \leq \beta \Bigl) \geq\tilde{\epsilon} \IEEEyessubnumber\label{eq:imcsi:2:g}\\
     		  &&\ \eqref{eq:8c}, \eqref{eq:imcsi:b} \IEEEyessubnumber\label{eq:imcsi:2:h}
\end{IEEEeqnarray}
where $\boldsymbol{\phi}=\{\phi_g\}$, $\boldsymbol{\alpha}=\{\alpha_l\}$, and $\beta$ are newly introduced variables. The constraint \eqref{eq:imcsi:2:e} is imposed to ensure that for a given CSI error set $\Omega_l$, the minimum received SINR at the $l$-th	PR is larger than the minimum SINR requirement $\alpha_l$ for the PR. According to \eqref{eq:imcsi:2:c} and \eqref{eq:imcsi:2:g}, the probabilities that the maximum received SINR at the $k_g$-th passive Eve and at the $k_p$-th Eve are  less than $\phi_{g} > 0$ and $\beta > 0$ are ensured to be greater than $\epsilon_{g}  $ and $\tilde{\epsilon} $, respectively.

 We are now in position to expose the hidden convexity of the constraint  of \eqref{eq:imcsi:2:c}, \eqref{eq:imcsi:2:e}, and \eqref{eq:imcsi:2:g}. Since $\mathbf{U}$ does not require a rank-constraint matrix, we introduce $ \widetilde{\mathbf{U}} \triangleq \mathbf{U}\mathbf{U}^H$ to facilitate the optimization problem.   Let us handle the constraint  \eqref{eq:imcsi:2:e} first by rewriting  as 
\begin{IEEEeqnarray}{rCl}\label{eq:imcsi:2:e:1}
\max_{\Delta\mathbf{f}_l\in\Omega_l}\sum\nolimits_{g=1}^G|\mathbf{f}_l^{H}\mathbf{w}_g|^2+ \tr(\mathbf{f}_l^{H} \widetilde{\mathbf{U}}\mathbf{f}_l)+\sigma_l^2 \leq \frac{P_p|h_l|^2}{\alpha_l}.
\end{IEEEeqnarray} 
For arbitrary $l$-th PR, \eqref{eq:imcsi:2:e:1} can be shaped to take the following equivalent form
\begin{IEEEeqnarray}{rCl}\label{eq:imcsi:2:e:2}
\sum_{g=1}^G\mu_{l,g} + \tilde{\mu}_l +\sigma_l^2 &\leq& \frac{P_p|h_l|^2}{\alpha_l}\label{eq:imcsi:2:e:2a}, l\in\mathcal{L}\\
\max_{\Delta\mathbf{f}_l\in\Omega_l}|\mathbf{f}_l^{H}\mathbf{w}_g|^2 &\leq&  \mu_{l,g},  l\in\mathcal{L}, g\in\mathcal{G}\label{eq:imcsi:2:e:2b}\\
\max_{\Delta\mathbf{f}_l\in\Omega_l} \tr(\mathbf{f}_l^{H} \widetilde{\mathbf{U}}\mathbf{f}_l) &\leq & \tilde{\mu}_l, l\in\mathcal{L}\label{eq:imcsi:2:e:2c}
\end{IEEEeqnarray} 
where $\boldsymbol{\mu}_l=\{\mu_{l,g}\}$ and $\boldsymbol{\tilde{\mu}}=\{\tilde{\mu}_l\}$ are new variables. Note that both sides of \eqref{eq:imcsi:2:e:2a} are convex, so it is iteratively replaced by the following linear constraint
\begin{IEEEeqnarray}{rCl}
\sum_{g=1}^G\mu_{l,g} + \tilde{\mu}_l +\sigma_l^2 &\leq& \frac{2P_p|h_l|^2}{\alpha_l^{(n)}} - \frac{P_p|h_l|^2}{(\alpha_l^{(n)})^2}\alpha_l,\label{eq:imcsi:2:e:2a1}l\in\mathcal{L}.
\end{IEEEeqnarray}
 To make the tractable form of \eqref{eq:imcsi:2:e:2b} and \eqref{eq:imcsi:2:e:2c}, we first transform these constraints into a  matrix inequality.
Substituting $\mathbf{f}_l=\mathbf{\hat{f}}_l+\Delta\mathbf{f}_l, \forall l$ into \eqref{eq:imcsi:2:e:2b} and applying \textit{S-Procedure} \cite{Stephen}, then        
\begin{equation}
\begin{aligned}
            &\Delta\mathbf{f}_l^{H}\Delta\mathbf{f}_l-\delta_l^2\leq0\\
\Rightarrow \eqref{eq:imcsi:2:e:2b}:\;&\Delta\mathbf{f}_l^{H}\mathbf{w}_g\mathbf{w}_g^H\Delta\mathbf{f}_l+2\Re\{\mathbf{\hat{f}}_l^{H}\mathbf{w}_g\mathbf{w}_g^H\Delta\mathbf{f}_l\}\\
&\qquad\qquad\qquad\quad+\mathbf{\hat{f}}_l^{H}\mathbf{w}_g\mathbf{w}_g^H\mathbf{\hat{f}}_l-\mu_{l,g}\leq 0 
\end{aligned}\end{equation}
holds if and only if there exists $\boldsymbol{\omega}_{l}=\{\omega_{l,g}\geq 0\},\;\forall l$, so that the matrix inequality constraint holds as 
\begin{equation}\begin{aligned}
\begin{bmatrix}
    \omega_{l,g}\mathbf{I}_N-\mathbf{w}_g\mathbf{w}_g^H & -\mathbf{w}_g\mathbf{w}_g^H\mathbf{\hat{f}}_l\\
    -\mathbf{\hat{f}}_l^{H}\mathbf{w}_g\mathbf{w}_g^H & -\mathbf{\hat{f}}_l^{H}\mathbf{w}_g\mathbf{w}_g^H\mathbf{\hat{f}}_l-\omega_{l,g}\delta_l^2+\mu_{l,g}
 \end{bmatrix}
\succeq\mathbf{0}.
\end{aligned}\label{eq:LMI:constraint:2}\end{equation}
However, \eqref{eq:LMI:constraint:2} is still not in a tractable form. At this point, we apply the application of Schur's complement lemma \cite[Eq.~(7.2.6)]{Tal:09} to obtain the following linear matrix inequality (LMI)
\begin{equation}\begin{aligned}
&\exists \omega_{l,g} \geq 0: \mathbf{C}_{l,g}(\mathbf{w}_g,\mu_{l,g},\omega_{l,g})\triangleq\\
&\begin{bmatrix}
    1 &     \mathbf{w}_g^H      &-\mathbf{w}_g^H\mathbf{\hat{f}}_l\\
		\mathbf{w}_g                                                    &  \omega_{l,g}\mathbf{I}_N  &                                                    \\
    -\mathbf{\hat{f}}_l^{H}\mathbf{w}_g & &-\omega_{l,g}\delta_l^2+\mu_{l,g}
 \end{bmatrix}
\succeq\mathbf{0},\ g\in\mathcal{G}, l\in\mathcal{L}.
\end{aligned}\label{eq:LMI:constraint:3}\end{equation}
It is also worth noting that constraint \eqref{eq:LMI:constraint:3} now includes only a finite number of constraints. 

Analogously, with $\tilde{\boldsymbol{\omega}}=\{\tilde{\omega}_{l}\geq 0\},\;$ the constraint \eqref{eq:imcsi:2:e:2c} admits the following representation
\begin{equation}\begin{aligned}
&\exists \tilde{\omega}_l \geq 0: \tilde{\mathbf{C}}_{l}(\widetilde{\mathbf{U}},\tilde{\mu}_{l},\tilde{\omega}_{l})\\
&\triangleq\begin{bmatrix}
    \tilde{\omega}_{l}\mathbf{I}_N - \widetilde{\mathbf{U}} &- \widetilde{\mathbf{U}}\mathbf{\hat{f}}_l\\
		-\mathbf{\hat{f}}_l^{H} \widetilde{\mathbf{U}} &-\mathbf{\hat{f}}_l^{H} \widetilde{\mathbf{U}}\mathbf{\hat{f}}_l-\tilde{\omega}_{l}\delta_l^2+\tilde{\mu}_{l}
 \end{bmatrix}
\succeq\mathbf{0},\ l\in\mathcal{L}.
\end{aligned}\label{eq:LMI:constraint:4}\end{equation}

To deal with the nonconvex constraints given in  \eqref{eq:imcsi:2:g} and \eqref{eq:imcsi:2:c}, we provide the following two lemmas, whose proofs are omitted due to space limitations. 
\begin{lemma}
For the primary system, the constraint in \eqref{eq:imcsi:2:g} is lower bounded by the following constraint
\begin{equation}
\lambda_{\min}\left(\sum\nolimits_{g=1}^G\mathbf{w}_g\mathbf{w}_g^H +\widetilde{\mathbf{U}}\right) \geq \tilde{\xi}(\beta)
\label{eq:lemma:2}\end{equation}
where $\tilde{\xi}(\beta) \triangleq  \Bigl(\exp\bigl(-\frac{\beta}{NP_p}\sigma_{k_p}^2\bigr)/(1-\tilde{\epsilon}^{1/K_p})^{1/N}-1\Bigl)\frac{P_p}{\beta}$ and $\lambda_{\min}(\mathbf{X})$ denotes the minimum eigenvalue of matrix $\mathbf{X}$.
\end{lemma}

Next, we rewrite \eqref{eq:lemma:2} equivalently in the form of
\begin{IEEEeqnarray}{rCl}
2\ln\eta + \beta\frac{\sigma_{k_p}^2}{NP_p} &\geq & 0 \label{eq:lemma2:a}\\
\bigl(\eta^2/(1-\tilde{\epsilon}^{1/K_p})^{1/N} -1\bigr)P_p &\leq& \beta \theta  \label{eq:lemma2:b}\\
\lambda_{\min}\Bigr(\sum\nolimits_{g=1}^G\mathbf{w}_g\mathbf{w}_g^H +\widetilde{\mathbf{U}}\Bigl) &\geq& \theta \label{eq:lemma2:c}
\end{IEEEeqnarray}
where $\theta$ and $\eta$ are newly introduced variables. We now focus on the nonconvex constraint. 
For the nonconvex constraint \eqref{eq:lemma2:c}, we note that both $\sum\nolimits_{g=1}^G\mathbf{w}_g\mathbf{w}_g^H$ and $\widetilde{\mathbf{U}}$ are Hermitian matrices. In addition,  the eigenvalues of a Hermitian matrix $\mathbf{Q}$ are real and satisfy $\tr(\mathbf{x}^H\mathbf{Q}^H\mathbf{x}) \ge \lambda\|\mathbf{x}\|^2$ for any given vector $\mathbf{x}$ if and only if $\lambda_{\min}(\mathbf{Q}) \geq \lambda$. Since $\lambda_{\min}(\mathbf{w}_g\mathbf{w}_g^H)=0$ for all $g$, the lower bound of right side of \eqref{eq:lemma2:c} is given by
\begin{IEEEeqnarray}{rCl}
\lambda_{\min}\Bigr(\sum\nolimits_{g=1}^G\mathbf{w}_g\mathbf{w}_g^H +\widetilde{\mathbf{U}}\Bigl)\geq \lambda_{\min}(\widetilde{\mathbf{U}}). \label{eq:lemma2:c1}
\end{IEEEeqnarray}
  From \eqref{eq:lemma2:c}, it follows that 
\begin{IEEEeqnarray}{rCl}
 \lambda_{\min}(\widetilde{\mathbf{U}}) \geq \theta \Leftrightarrow \widetilde{\mathbf{U}} \succeq \mathbf{I}_N\theta \label{eq:lemma2:c2}.
\end{IEEEeqnarray}

\begin{lemma}
For the secondary system, the constraint in \eqref{eq:imcsi:2:c}  is lower bounded by the following constraint
\begin{equation}
\frac{\|\mathbf{w}_g\|^2}{\phi_g} \leq \xi_g  + \sum\nolimits_{i=1,i\neq g}^G\|\mathbf{w}_i\|^2 + \lambda_{\min}(\widetilde{\mathbf{U}}),\ g\in\mathcal{G}
\label{eq:lemma:3}\end{equation}
where $\xi_g\triangleq\Bigl[\exp\Bigl(\frac{\sigma_{k_g}^2}{NP_p}\Bigr)\epsilon_{g}^{-1/NK_g}-1\Bigr]P_p$.
\end{lemma}

The formulation in  \eqref{eq:lemma:3} can be further shaped to take the following convex constraints
\begin{IEEEeqnarray}{rCl}
&&\frac{\|\mathbf{w}_g\|^2}{\phi_g} \leq \xi_g  + \sum\nolimits_{i=1,i\neq g}^G 2\Re\{(\mathbf{w}_i^{(n)})^H\mathbf{w}_i\} \nonumber\\
&&\qquad\qquad\qquad - \sum\nolimits_{i=1,i\neq g}^G\|\mathbf{w}_i^{(n)}\|^2 + \vartheta,\ g\in\mathcal{G} \label{eq:lemma:3a}\\
&&\lambda_{\min}(\widetilde{\mathbf{U}}) \geq \vartheta \Leftrightarrow  \widetilde{\mathbf{U}} \succeq \mathbf{I}_N \vartheta \label{eq:lemma:3c}
\end{IEEEeqnarray}
where $\vartheta$ is newly introduced variable.

With the above discussions, the approximate convex problem solved at  $(n+1)$-th iteration of the  proposed design is given by
\begin{IEEEeqnarray}{rCl}\label{eq:imcsi:3}
&&\underset{\substack{\mathbf{w}, \widetilde{\mathbf{U}}\succeq\mathbf{0}, \boldsymbol{t}, z, \varphi, \boldsymbol{\phi},\boldsymbol{\alpha},\\ \beta, \boldsymbol{\mu}_l,\tilde{\boldsymbol{\mu}}, \boldsymbol{\omega}_l,\tilde{\boldsymbol{\omega}},\theta, \eta, \vartheta}}{\mathrm{\mathrm{maximize}}}\quad  \varphi \IEEEyessubnumber\\
  && \st \quad   
	 \mathcal{F}_{m_g}^{(n)}(\mathbf{w},\widetilde{\mathbf{U}}) \geq (\varphi + t_g)\ln2,\ m_g\in\mathcal{S}_{g},g\in\mathcal{G} \quad\IEEEyessubnumber \\
     		  &&\quad  \sum\nolimits_{g=1}^G\|\mathbf{w}_g\|^2+ \tr(\widetilde{\mathbf{U}}) \leq P_{s}\, \IEEEyessubnumber\label{eq:8c:revised} \\
					&&\quad   \eqref{eq:imcsi:2:b}, \eqref{eq:imcsi:2:d}, \eqref{eq:imcsi:2:f}, \eqref{eq:imcsi:2:e:2a1}, \eqref{eq:LMI:constraint:3},  \nonumber \\
					&&\quad \eqref{eq:LMI:constraint:4}, \eqref{eq:lemma2:a}, \eqref{eq:lemma2:b}, \eqref{eq:lemma2:c2}, \eqref{eq:lemma:3a},  \eqref{eq:lemma:3c}. \IEEEyessubnumber\label{eq:imcsi:3:h}
\end{IEEEeqnarray}
To find an initial feasible point to \eqref{eq:imcsi:1}, we solve the following convex optimization problem
\begin{IEEEeqnarray}{rCl}\label{ipcsi:ini3.m}
&&\max_{\substack{\mathbf{w}, \widetilde{\mathbf{U}}\succeq\mathbf{0},  z,\boldsymbol{\alpha}, \beta,\\   \boldsymbol{\mu}_l,\tilde{\boldsymbol{\mu}}, \boldsymbol{\omega}_l,\tilde{\boldsymbol{\omega}},\theta,\eta }}\;\min_{l\in\mathcal{L}}\;
\Bigl\{\log_2\bigl(1+\alpha_l\bigl) - z - \bar{R}_{p,l}
  \Bigr\}\IEEEyessubnumber \label{eq:ipcsi:ini3:a}\\
	&&\ \st\ \,  \eqref{eq:imcsi:2:f}, \eqref{eq:imcsi:2:e:2a1}, \eqref{eq:LMI:constraint:3}, \eqref{eq:LMI:constraint:4},\eqref{eq:lemma2:a}, \eqref{eq:lemma2:b}, \eqref{eq:lemma2:c2}, \eqref{eq:8c:revised} \IEEEyessubnumber \label{eq:ipcsi:ini3:c}
\end{IEEEeqnarray}
and  stop at reaching: $\min_{l\in\mathcal{L}}\; \Bigl\{\log_2\bigl(1+\alpha_l\bigl) - z - \bar{R}_{p,l}\Bigr\}\geq 0.$

The proposed iterative method is outlined in Algorithm \ref{algo:proposed:DUAL:2}. We can show  that Algorithm \ref{algo:proposed:DUAL:2} yields a nondecreasing  sequence of the objective value due to  updating the involved variables after each iteration, which converges to a KKT point \cite{Marks:78}.
\begin{algorithm}[t]
\begin{algorithmic}[1]

\protect\caption{The proposed iterative algorithm to solve \eqref{eq:imcsi:1}}

\label{algo:proposed:DUAL:2}

\global\long\def\algorithmicrequire{\textbf{Initialization:}}

\REQUIRE  Set $n:=0$ and solve \eqref{ipcsi:ini3.m} to generate an  initial feasible point $\bigl(\mathbf{w}^{(n)},\widetilde{\mathbf{U}}^{(n)}, \boldsymbol{\alpha}^{(n)}\bigr)$ 

\REPEAT
\STATE Solve \eqref{eq:imcsi:3} to obtain the optimal solution: $\bigl(\mathbf{w}^{*},\widetilde{\mathbf{U}}^{*}, \boldsymbol{\alpha}^{*})$.

\STATE Update\ $\mathbf{w}^{(n+1)}:=\mathbf{w}^{*}$,  $\widetilde{\mathbf{U}}^{(n+1)}:=\widetilde{\mathbf{U}}^{*}$, and $\boldsymbol{\alpha}^{(n+1)}:=\boldsymbol{\alpha}^{*}$.

\STATE Set $n:=n+1.$
\UNTIL Convergence or maximum required number of iterations\\
\end{algorithmic} \end{algorithm}

\textit{Complexity Analysis}: The optimization problem in \eqref{eq:imcsi:3}  involves $GL$ LMI constraints of size $N+2$, $L$ LMI constraints of size $N+1$, and 2 LMI constraints of size $N$. In each iteration of Algorithm 1, the worst-case computational complexity for solving the generic convex problem in \eqref{eq:imcsi:3} using interior point methods is given by $\mathcal{O}\Bigl(n\sqrt{GL(N+2) + L(N+1) + 2N}\bigl[GL(N+2)^3 + L(N+1)^3 + 2N^3 + nGL(N+2)^2 + nL(N+1)^2 + 2nN^2 +n^2\bigl]\Bigl)$, where $n = G(L+3)+N(N+G)+2L+6$ \cite{Ben:2001}.

\vspace{-0.3cm}
\section{Numerical Results And Discussions}\label{Numerical}
The number of groups of SUs is set to $G=2$,  each of which consists of two SR users, i.e., $M_g=2,\,\forall g$. The number of  PR is set to $L=2$, and each group of SUs and PUs is surrounded by two Eves, i.e., $K_p = K_g = 2$. All  channel entries are assumed to be i.i.d. complex Gaussian random variables with $\mathcal{CN}(0,1)$, and the background thermal noise at each user is generated as i.i.d. complex Gaussian random variables  with zero means and unit variance. The transmit power at the PT is fixed to $P_p=20$ dBm. For simplicity, we further assume that the minimum secrecy rate requirement for all PUs are the same, i.e., $\bar{R}_{p,l}=\bar{R}_{p}, \forall l$.  For the imperfect CSI of the PU channels, we define the normalized channel estimation errors as $\bar{\delta}^2_l=\delta^2_l/\|\mathbf{f}_l\|^2 =5\%$, $\forall l$. To guarantee secure communications, we choose $\tilde{\epsilon}=0.99$ and $\epsilon_g=0.99,\,\forall g$ for the passive Eves.

\begin{figure}
           \includegraphics[width=0.45\textwidth,trim={-1.0cm 0cm 0.5cm -0.0cm}]{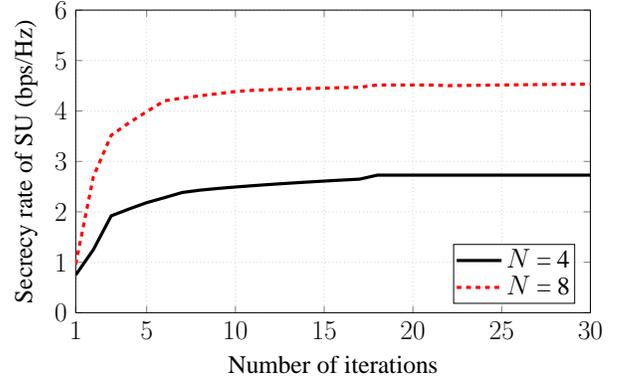}
    			  \caption{Convergence results of Algorithm 1  for different numbers of antennas at the ST over one random channel realization with $\bar{R}_{p} = 2$ bps/Hz and $P_s$ = 15 dBm.}\label{fig:Convergencebehavior:Iteration}
\end{figure}

Fig.~\ref{fig:Convergencebehavior:Iteration} illustrates the typical convergence behavior of the proposed Algorithm 1. As seen, the objective value of the proposed algorithm increases rapidly within the first 10 iterations and stabilize  after a few more iterations, and its convergence rate is slightly sensitive to the problem size i.e., as $N$  increases. The convergence results also confirm that all optimization variables are accounted  to find a better solution for the next iteration, i.e., the secrecy rates of SUs  monotonically increasing. 

\begin{figure}[t]
\centering
\includegraphics[width=0.385\textwidth,trim={0.2cm 0.0cm 0.2cm 0.0cm}]{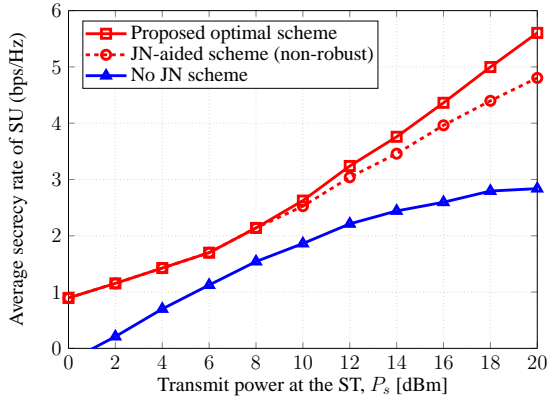}
\caption{ Average secrecy rate of the secondary system  vs. the transmit power at the ST, where $\bar{R}_{p} = 1$ bps/Hz and $N = 8$.}
\label{fig:Fig6:Ps:IPCSI}
\end{figure}

We  compare the performance of the proposed scheme with  the known solutions, namely the ``No JN scheme'' \cite{Pei} and ``JN-aided scheme (non-robust)''. In ``No JN scheme'', we set $\mathbf{U}$  to $\mathbf{0}$.  For the non-robust secrecy rate design, we use the presumed CSIs as $\hat{\mathbf{f}}_l,\ \forall l$ rather than the true ones, to perform the transmit design, which then evaluates the resultant  secrecy rate. Fig.~\ref{fig:Fig6:Ps:IPCSI} depicts the  secrecy rate  as a function of the transmit power at the ST. As can be observed that the secrecy rate of non-robust design  is sensitive to the CSI uncertainties for high $P_s$. In particular, when $P_{s} \geq 8$ dBm, the non-robust design exhibits the degradation in terms of the secrecy rate that tends to worsen as $P_{s} $ increases. Moreover, the proposed optimal design achieves the best secrecy rate performance, compared to the other designs.

\begin{figure}
           \includegraphics[width=0.394\textwidth,trim={-1.5cm .00cm 1.5cm 0.0cm}]{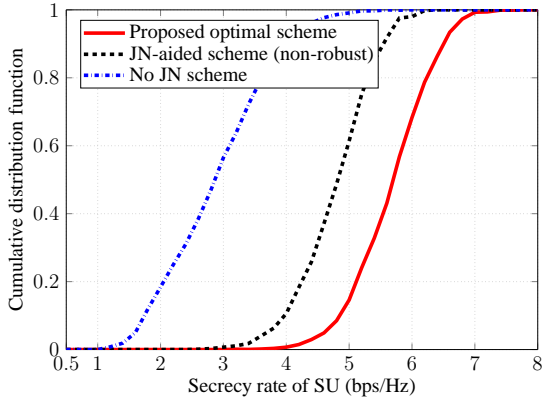}
 	  \caption{CDF of secrecy rate of the secondary system for  different schemes, where $\bar{R}_{p} = 1$ bps/Hz, $N$ = 8, and $P_s$ = 20 dBm.}\label{fig:CDF}
\end{figure}

Finally, we generate cumulative distribution function (CDF) of the secrecy rate of the secondary system  in Fig.~\ref{fig:CDF} for different schemes. It is obvious in  CDF that on account for a larger feasible set, the proposed optimal scheme can promise a bigger secrecy rate as expected.  For instance, the proposed optimal scheme attains 0.8 bps/Hz and 2.8 bps/Hz of the achievable secrecy rate  higher than the non-robust scheme and ``No JN scheme'', respectively, for approximately $60\%$ of the simulated trials.

\vspace{-0.3cm}
\section{Conclusion}\label{Conclusion}
In this paper, we  have proposed  PHY security for both  primary and secondary systems in the presence of the multiple secondary receiver groups and multiple primary receivers.  The main objective  is to maximize the secrecy rate of the secondary system, while the secondary transmitter is constrained not only by  power constraint, but also by the individual the minimum secrecy rate requirements of the primary users.  We have proposed iterative  algorithms to solve the optimization problem based on a convex formulation in each iteration.  We have carried out  simulation  to evaluate the advantages of the proposed design. 

\vspace{-0.15cm}
\section*{Acknowledgment}
This work was supported in part by the U.K. Royal Academy of Engineering Research Fellowship under Grant RF1415$\backslash$14$\backslash$22 and U.K. Engineering and Physical Sciences Research Council under Grant EP/P019374/1.

\end{document}